\documentclass{osa-article}

\journal{oe}
\articletype{Research Article}
\begin{document}

\title{Circularly polarized extreme ultraviolet high harmonic generation in graphene}

\author{Zi-Yu Chen\authormark{1,2} and Rui Qin\authormark{1,3}}

\address{\authormark{1}National Key Laboratory of Shock Wave and Detonation Physics, Institute of Fluid Physics, China Academy of Engineering Physics, Mianyang 621999, China\\
\authormark{2}ziyuch@caep.ac.cn\\
\authormark{3}qinrui.phy@outlook.com
}

%\email{\authormark{*}ziyuch@caep.ac.cn} %% email address is required



\begin{abstract}
Circularly polarized extreme ultraviolet (XUV) radiation is highly interesting for investigation of chirality-sensitive light-matter interactions. Recent breakthroughs have enabled generation of such light sources via high harmonic generation (HHG) from rare gases. There is a growing interest in extending HHG medium from gases to solids, especially to 2D materials, as they hold great promise to develop ultra-compact solid-state photonic devices and provide insights into electronic properties of the materials themselves. However, HHG in graphene driven by terahertz to mid-infrared fields reported so far only generate low harmonic orders, and furthermore no harmonics driven by circularly polarized lasers. Here, using first-principles simulations within a time-dependent density-functional theory framework, we show that it is possible to generate HHG extending to the XUV spectral region in monolayer extended graphene excited by near-infrared lasers. Moreover, we demonstrate that a single circularly polarized driver is enough to ensure HHG in graphene with circular polarization. The corresponding spectra reflect the six-fold rotational symmetry of the graphene crystal. Extending HHG in graphene to the XUV spectral regime and realizing circular polarization represent an important step towards the development of novel nanoscale attosecond photonic devices and numerous applications such as spectroscopic investigation and nanoscale imaging of ultrafast chiral and spin dynamics in graphene and other 2D materials.
\end{abstract}

%%%%%%%%%%%%%%%%%%%%%%%%%%  body  %%%%%%%%%%%%%%%%%%%%%%%%%%
\section{Introduction}
Circularly polarized radiation source in the extreme ultraviolet (XUV) spectral regions is a unique and powerful tool to probe chirality-sensitive light-matter interactions, from photoionization in chiral molecules\cite{Garcia2013} to magnetic properties in solid-state materials\cite{Schmising2014}. There has been a quest for generating such light sources on the table-top scale. Table-top XUV source is generally produced through high harmonic generation (HHG), an extremely nonlinear optical process up-converting driving laser frequency, in rare gases. However, it had been believed circularly polarized high harmonics could not be generated for many years, because electron trajectory undergoes lateral drifting in a circularly polarized driving field so that the electron cannot recollide with its parent ion. In the past few years, intensive efforts have been made towards circularly polarized HHG. Use of two counter-rotating circularly polarized laser fields has lead to the breakthrough of circularly polarized high harmonic\cite{Ferre2015,Kfir2015,Hickstein2015} and isolated attosceond pulse\cite{Huang2018} generation experimentally. Later, it was numerically demonstrated that circularly polarized XUV and attosceond pulses can also be generated via HHG from relativistic plasma surfaces\cite{Chen2016,Ma2016,Chen2018a,Chen2018b}.

Recently, there has been a growing interest to extend HHG and related techniques to solid-state systems. HHG in solids not only offers a novel approach to XUV and attosecond photonics, but also provides a new platform to study structure and ultrafast strong-field dynamics in the condensed phase. XUV HHG has been generated in bulk crystals\cite{Ghimire2011,Schubert2014,Vampa2015a,Ndabashimiye2016,You2017} and demonstrated useful for probing electronic properties, such as reconstruction of bandgap\cite{Vampa2015b}, retrieving energy dispersion profile of conduction band\cite{Luu2015}, and measurement of Berry curvature\cite{Luu2018}. In contrast to atomic HHG, HHG driven by a single-color circularly polarized laser field is shown to be possible in bulk solids. Generation of circularly polarized XUV HHG has been predicted in cubic Si and MgO crystals\cite{TD2017b}.

Apart from bulk crystals, nonperturbative HHG in two-dimensional (2D) materials has also attracted much attention\cite{Liu2017,TD2018}, as they exhibit distinct electronic properties compared to the bulk. In particular, HHG in graphene, the most popular and promising 2D material with zero band gap and massless Dirac fermions, has been actively studied both theoretically\cite{Mikhailov2007,Mikhailov2008,Ishikawa2010,Avetissian2012a,Avetissian2012b,AN2014,AN2015,
Chizhova2016,Chizhova2017,Cox2017,Avetissian2018,cdLiu2018} and experimentally\cite{Bowlan2014,Yoshikawa2017,Taucer2017,Baudisch2018}. With linear energy dispersion, graphene is expected to display strong nonlinear optical responses. However, the highest harmonic order observed in extended undoped graphene so far has been limited to less than the 10th order driven by laser wavelength from terahertz (THz) to mid-infrared\cite{Bowlan2014,Yoshikawa2017,Taucer2017,Baudisch2018}, which hampers the possibility of developing graphene-based compact XUV sources. In addition, several groups have studied ellipticity dependence of HHG in graphene yet come out with conflicting results. Yoshikawa \textit{et al.} show that HHG can be enhanced at a small non-zero laser ellipticity\cite{Yoshikawa2017}, while Taucer \textit{et al.}\cite{Taucer2017} and Baudisch \textit{et al.}\cite{Baudisch2018} present an atomic-HHG-like monotonic ellipticity dependence. Moreover, the reported ellipticity dependencies suggest that HHG is greatly suppressed by using circularly polarized driving field. These lead one to question whether circularly polarized XUV high harmonics can be generated in graphene or not.

In this work, we perform, to our knowledge, the first \textit{ab initio} calculations of HHG in monolayer graphene based on the framework of time-dependent density-functional theory (TDDFT)\cite{Runge1984,Leeuwen1998,Castro2004a}. Employing more intense light pulses at a higher photon frequency (1.55 eV), simulation results show high harmonics up to the 15th order can be obtained with a pump intensity of 3 TW/cm$^2$ and pulse duration of 15 fs, and thus HHG in graphene can be extended to the XUV spectral regions. We then study ellipticity dependence of HHG in graphene. The results show that it is possible to generate high harmonics in graphene with a single circularly polarized driving pulse. The corresponding spectra reflect the six-fold rotational symmetry of the graphene crystal. Moreover, we demonstrate the generated high harmonics are also circularly polarized. These predictions show the potential of circularly polarized XUV HHG in 2D materials as novel ultra-compact XUV photonic sources and in enabling nanoscale imaging and spectroscopic investigation of spin, magnetic, and other chirality-related phenomena in 2D materials.

\section{Methods}

Graphene structures are studied by using the semiperiodic supercell model, where a hexagonal primitive cell contains two carbon atoms. We optimize the graphene structure. The C-C bond length is found to be 1.413 \AA. Vacuum space of 30 Bohr, including 3 Bohr of absorbing regions on each side of the monolayer, is chosen and well tested to simulate the single-layer graphene in a supercell and avoid reflection in the spectral region of interest. The period of the reciprocal space of graphene along the armchair (AC) and zigzag (ZZ) directions is 2.964 \AA$^{-1}$ and 5.135 \AA$^{-1}$, respectively. The Octopus package\cite{Andrade2015,Castro2006,Andrade2012} is employed to perform the simulations. The ground state electronic structures and geometric structure relaxation are performed within the density functional theory (DFT) framework in the local density approximation (LDA)\cite{Marques2012}. Time evolution of the wave functions and time-dependent electronic current are calculated by propagating the Kohn-Sham equations in real time and real space\cite{Castro2004b} within the TDDFT framework in the adiabatic LDA (ALDA). No dephasing term is added in the calculations. The real-space spacing is 0.4 Bohr. A $60\times 60\times 1$ Monkhorst-Pack \textit{k}-point mesh for the BZ sampling is used, and the sampling is scaled according to the size of the supercells. The fully relativistic Hartwigsen, Goedecker, and Hutter (HGH) pseudopotentials are adopted. All the simulations were performed using the supercomputers at the National Supercomputing Center in Shenzen, China. Each run took about 4 hours using 1024 paralleled processors (AMD 6136).

Laser field is described in the velocity gauge. The vector potential can be expressed as\cite{TD2017b}
\begin{equation}
\textbf{A}(t)=\frac{\sqrt{I_L c}}{\omega}f(t)\Big[ \frac{1}{\sqrt{1+\epsilon^2}} \cos(\omega t +\phi)\hat{\textbf{e}}_x + \frac{\epsilon}{\sqrt{1+\epsilon^2}} \sin(\omega t +\phi)\hat{\textbf{e}}_y \Big],
\end{equation}
where $I_L$ is the laser peak intensity inside matter, $\omega$ is the laser photon frequency, $\phi$ is the carrier-envelop phase (CEP), $f(t)$ is the pulse envelope, $\epsilon$ is the laser ellipticity, $c$ is the light speed in vacuum, and $\hat{\textbf{e}}_x$ and $\hat{\textbf{e}}_y$ are the unit vectors.
We use Ti:sapphire laser pulse with a wavelength of $\lambda_L=$ 800 nm (corresponding to a photon energy of 1.55 eV) and FWHM pulse duration of $\tau$ = 15 fs. The pulse envelope profile is sin-squared and the CEP is taken to be $\phi= 0$. The peak laser intensity is in the range of $I_L=1\times10^{12}$ W/cm$^{2}$ and $I_L=3\times10^{12}$ W/cm$^{2}$. The laser field is normally incident onto the graphene sample so that the driving electric field is in the plane of the monolayer.

The HHG spectra are calculated from the time-dependent electronic current density $\textbf{j}(\textbf{r},t)$ as
\begin{equation}
\mathrm{HHG}(\omega) = \Big| \mathcal{FT} \Big(\frac{\partial}{\partial t} \int \textbf{j}(\textbf{r},t) \ \mathrm{d}^3 \textbf{r}  \Big) \Big|^2,
\end{equation}
where $\mathcal{FT}$ denotes the Fourier transform. 

\section{Results}

\begin{figure}[htbp]
\centering
\includegraphics[width=0.95\textwidth
]{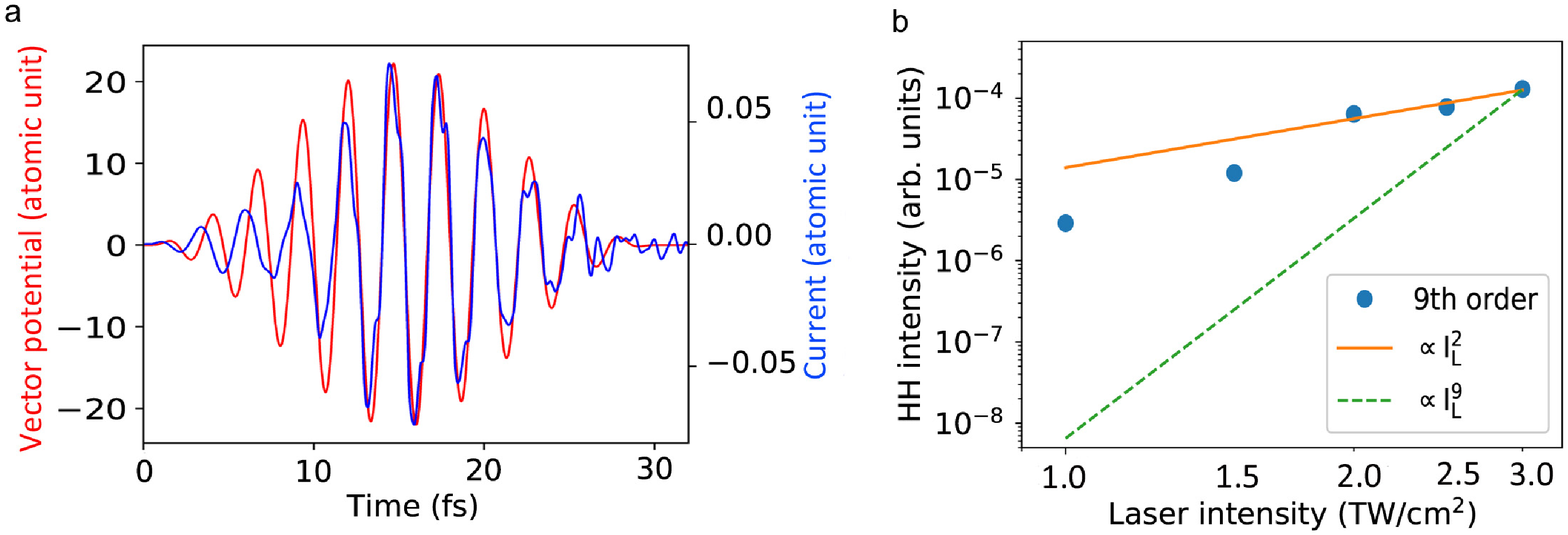}% Here is how to import EPS art
\caption{\label{nonpert} (a) Typical waveform of the applied vector potential (red) and induced electronic current (blue). The 800 nm near-infrared laser has a FWHM pulse duration of 15 fs and intensity of $I_L=3$ TW/cm$^2$. (b) Intensity of the 9th harmonic (blue dots) scales as $I_L^2$ (orange line) instead of $I_L^9$ (dashed green line), showing nonperturbative characteristic of the higher-order harmonic generation process studied here.}
\end{figure}

Typical waveforms of the applied vector potential (red line) and induced electronic current (blue line) are shown in Fig. \ref{nonpert}(a). The temporal evolution of the induced current tends to follow the driving laser profile. Distortion and nonlinear response in the current profile can also be seen, which lead to harmonic components in the radiation spectrum. Figure \ref{nonpert}(b) shows the intensity of 9th-harmonic radiation (blue dots) as a function of the driving laser intensity $I_L$ in the range of 1 TW/cm$^2$ and 3 TW/cm$^2$. The harmonic intensity scales as $I_L^2$ (orange line) at the highest pump intensity, in agreement with the previous studies\cite{Yoshikawa2017,Taucer2017}. This power law dependence clearly shows the higher-order harmonic generation process in this intensity range is in the nonperturbative regime, whereas it should behave as $I_L^9$ dependence (dashed green line) for the 9th harmonic in the perturbative limit.

\begin{figure}[htbp]
\centering
\includegraphics[width=0.7\textwidth
]{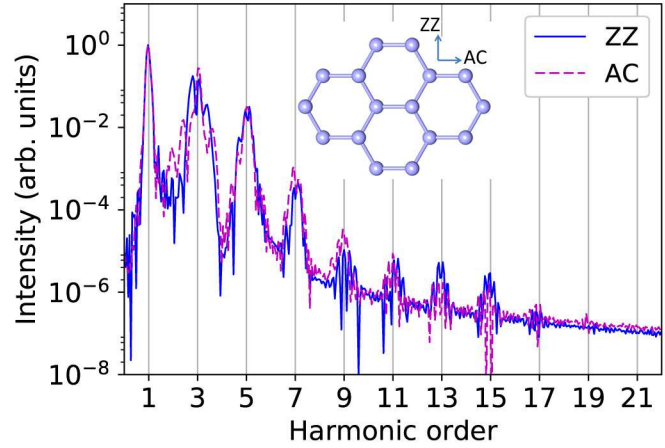}% Here is how to import EPS art
\caption{\label{spectra} High harmonic spectra up to the 15th order with a driving intensity of 3 TW/cm$^2$, demonstrating the possibility of HHG in graphene extending to the XUV spectral region. The linearly polarized fundamental field is directed along either the armchair (AC) or zigzag (ZZ) directions of the crystal. The inset shows the crystal structure of monolayer graphene and the definition of the AC and ZZ directions.}
\end{figure}

Figure \ref{spectra} shows the high harmonic spectra generated with a pump laser intensity of 3 TW/cm$^2$. Harmonics up to the 15th orders of the fundamental frequency (1.55 eV) are obtained, which already extend to the XUV spectral region. Only odd harmonic orders are present, reflecting the centrosymmetric nature preserved in the crystal lattice. Orienting the linearly polarized fundamental field along the AC and ZZ directions of the crystal result in different HHG spectra. The difference in peak harmonic intensity is more evident for higher-order harmonics. Therefore, graphene, though with isotropic Dirac cones, can give rise to anisotropic emission of high harmonics. This is because the graphene band structure is isotropic only in the vicinity of the Dirac points while exhibits strong anisotropy away from the Dirac points, in agreement with previous discussions\cite{Qin2018}.

It should be noted that previous experiments have reported a single-shot damage threshold of graphene to be $\sim$3 TW/cm$^2$ with a 790 nm and 50 fs laser pulse excitation\cite{Roberts2011}. In contrast, the pulse duration in our simulations is much shorter, i.e., only 15 fs. For the same intensity, a shorter pulse duration means less energy contained in the pulse, and thus less energy deposited in the material. Then the material is less likely to be damaged. Therefore, we can say that our simulations are performed below the single-shot damage threshold of the material. Yet, it is worth noting that even with intensity below the single-shot damage threshold, long-time multiple exposures can eventually lead to degradation of the lattice due to defect formation. However, providing a fresh target surface for each shot or a few shots can avoid or mitigate this long-term operation effect. In addition, compared to bulk crystals, graphene has the advantage of spooling, e.g., as a tape target\cite{Bierbach2015,Shaw2016}. Apart from the laser, we mention that samples used in real experiments usually contain defects and impurities, which should lower the damage threshold. Therefore, improving the sample quality is beneficial to high-order harmonic generation because even higher laser intensity may be applicable.  

\begin{figure}[htbp]
\centering
\includegraphics[width=0.95\textwidth
]{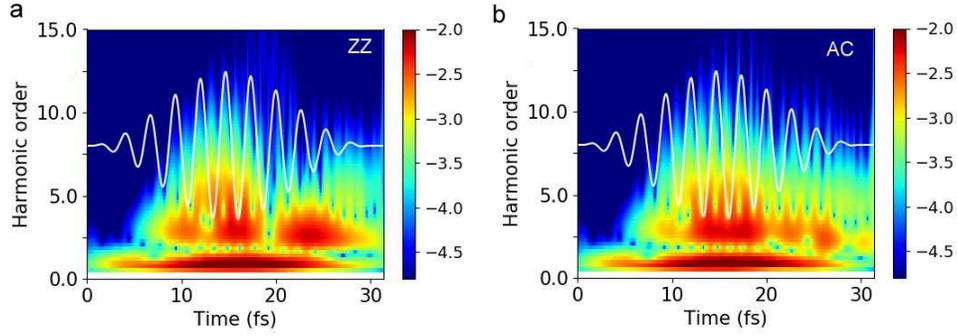}% Here is how to import EPS art
\caption{\label{wt} Evolution of the HHG spectra as a function of time, i.e., spectrogram of the HHG process, for laser polarization orientating along the (a) ZZ and (b) AC direction respectively, revealing the high harmonic emission in phase at each field peak, suggesting intraband contribution is the dominate mechanism for the HHG in this study. The white curves are the waveform of the laser pulses. The pump laser intensity is 3 TW/cm$^2$. Colorbar represents spectral intensity (arb. units) in logarithmic scale.}
\end{figure}

The mechanisms responsible for nonperturbative HHG in solids can be attributed to interband transition and intraband contribution. The former corresponds to direct electron-hole recombination, similar to the three-step mode of gas HHG, and thus characteristic recombination trajectories can be observed in the spectrogram\cite{Vampa2015c}, i.e., a 2D map of the emission in time and frequency. The latter mechanism corresponds to carriers accelerated within the energy bands driven by the laser field, and thus the harmonics are emitted in phase with each laser peak when the acceleration is maximum\cite{Vampa2017}. To gain insight into the physical dynamics underlying XUV emission in our calculations, we performed a time-frequency wavelet analysis of the harmonic emission. Figures \ref{wt}(a) and \ref{wt}(b) show the spectrogram with a pump intensity of 3 TW/cm$^2$ for laser polarization along the ZZ and AC directions, respectively. It clearly shows the high harmonics are emitted as discrete bursts in phase at each peak of the laser field corresponding to maximum electron acceleration. This in-phase signature in the spectrogram, instead of recombination trajectories, indicates intraband contribution being the dominate mechanism in the HHG process we studied here, in agreement with that from model calculations for HHG in monolayer graphene driven by a mid-infrared laser pulse\cite{Taucer2017}. We notes that the band structure of graphene near the Dirac cone has linear energy dispersion and is discontinuous at the Dirac point. At higher energy bands, the dispersion relation is also highly anharmonic. Then large intraband contribution is possible.

\begin{figure}[htbp]
\centering
\includegraphics[width=1.0\textwidth
]{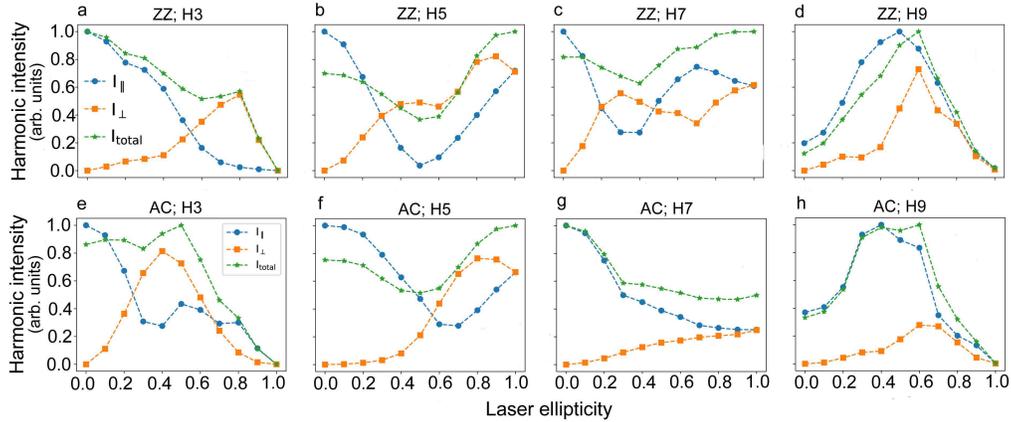}% Here is how to import EPS art
\caption{\label{ellip} Ellipticity dependence of the peak harmonic intensity for different harmonic orders. The major axis of the elliptical polarization of the driving field is fixed relative to the (a-d) ZZ and (e-h) AC directions. The driving field intensity is 3 TW/cm$^2$. For each harmonic order, intensity of the field components parallel ($I_{\parallel}$) or perpendicular ($I_{\perp}$) to the major axis of the laser polarization is normalized by the maximum intensity between the two (max$\{ I_{\parallel},I_{\perp} \}$). The total intensity ($I_{\text{total}}=I_{\parallel}+I_{\perp}$) is normalized by the maximum $I_{\text{total}}$. For the 5th and 7th harmonics, harmonic intensity is non-zero with circularly polarized driving pulse, in contrast to the 3rd and 9th harmonics.}
\end{figure}

Figure \ref{ellip} illustrates the laser-ellipticity dependence of the peak intensity of the 3rd, 5th, 7th, and 9th harmonic radiation. The major axis of the elliptical polarization of the driving laser is fixed relative to either the ZZ or the AC direction of the crystal. The incident pulse energy is constant while changing the laser ellipticity. For each harmonic order, the plotted intensity of the field components parallel ($I_{\parallel}$) or perpendicular ($I_{\perp}$) to the major axis of the laser polarization is normalized by the maximum intensity between the two (max$\{ I_{\parallel},I_{\perp} \}$); while the total intensity ($I_{\text{total}}=I_{\parallel}+I_{\perp}$) is normalized by the maximum $I_{\text{total}}$. HHG in graphene studied here displays complex ellipticity dependence. We observe different dependence trend for different harmonic orders. For the 3rd harmonics, $I_{\parallel}$ decreases gradually as increasing ellipticity, while $I_{\perp}$ is enhanced at a finite laser ellipticity and then drops. These features are consistent with the essential results reported by Yoshikawa \textit{et al.}\cite{Yoshikawa2017}. On the other hand, the monotonic decrease of $I_{\parallel}$ and $I_{\text{total}}$ with ellipticity is in agreement with the observations of Taucer \textit{et al.}\cite{Taucer2017} and Baudisch \textit{et al.}\cite{Baudisch2018}. The fact that different intensity is measured may explain the discrepancy between the previous results. Apart from the 3rd harmonic, other harmonic orders show distinct ellipticity dependence. For the 5th and 7th harmonics, $I_{\parallel}$ and $I_{\text{total}}$ firstly drop and then increase (except for the 7th harmonic at AC configuration where they decrease monotonically), while $I_{\perp}$ gradually increase with ellipticity. For the 9th harmonics, all the intensities display similar ellipticity dependence, i.e., gradually increasing to maximum values before decreasing close to zero. These results show the possibility of tuning and enhancing harmonic emission in graphene by using a finite ellipticity of the driving field. 

For the 5th and 7th harmonics, the harmonic yield $I_{\perp}$ for circularly polarized driver is higher than a linearly polarized driver. This is easily understandable, as the driving component in the perpendicular direction increases with the laser ellipticity. Yet, in some cases (see Figs.~\ref{ellip}(b), \ref{ellip}(c) and \ref{ellip}(f)), the overall harmonic yield for circularly polarized driver is still higher than a linearly polarized driver. This has not been reported before. We attribute this to the higher photon energy (1.55 eV) and higher laser intensity employed here. In this case, deeper electron bands are involved in the HHG process. Since the main HHG mechanism here is intraband contribution, the HHG is strongly affected by the potential energy landscape, i.e., the band dispersion or band structure, felt by the electrons driven by the strong laser field. As mentioned earlier, the band structure is isotropic in the vicinity of the Dirac cone but highly anisotropic away from the Dirac point. As a circularly polarized laser can drive electrons to explore the band dispersion along different directions, it may experience more anharmonicity than just oscillating along one direction driven by a linearly polarized driver. Therefore, it is possible to get higher harmonic yield for an intense circularly polarized driver. Previous experiments used laser with much lower photon energy and intensity. Those HHG process may be in a markedly different scenario, where only the bands near the Dirac cones are involved. Then, as illustrated by Baudisch \textit{et al.}\cite{Baudisch2018}, the charge carriers can re-encounter the lowest energy point twice per field cycle driven by linearly polarized fields, while exhibit spiraling trajectories with canceled anharmonic response in the Dirac potential driven by circularly polarized fields. Consequently, the harmonic yield is lower for circularly polarized driver in those studies.

\begin{figure}[htbp]
\centering
\includegraphics[width=1.0\textwidth
]{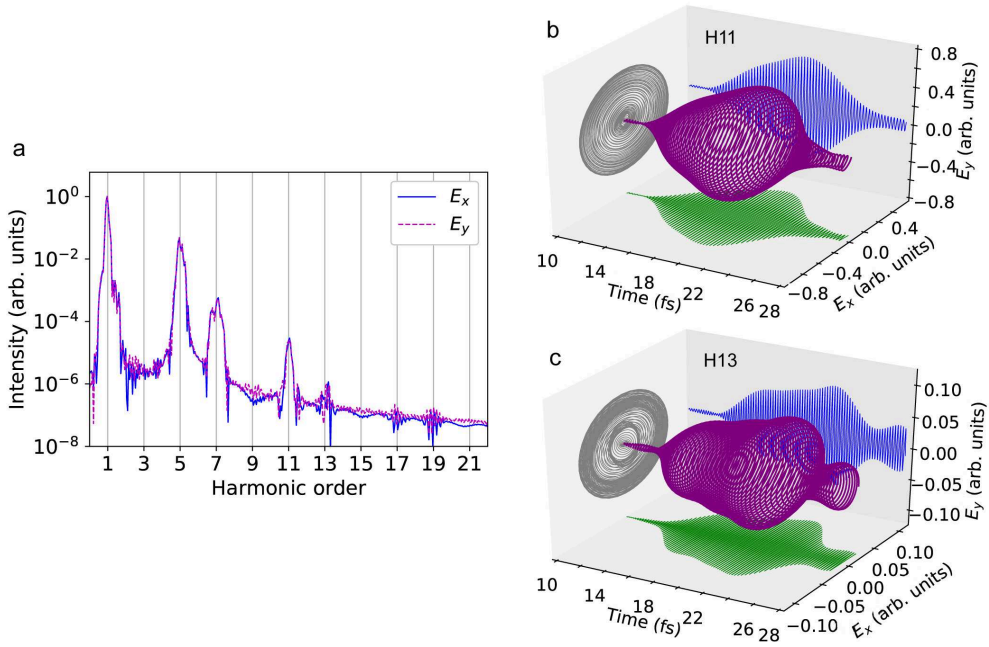}% Here is how to import EPS art
\caption{\label{cp} (a) High harmonic spectra in graphene driven by a single circularly polarized laser field with intensity of 3 TW/cm$^2$. The selection rule of the harmonic orders ($n = 6m \pm 1$ ($m = 1,2,3,\cdots$)) reflects the six-fold rotational symmetry of the graphene crystal. 3D plots of the electric field vector (purple) of the (b) 11th and (c) 13th harmonic demonstrate the generated high harmonics are circularly polarized with opposite helicity. Also shown in panels (b) and (c) are the waveform of the two orthogonal electric field components $E_x$ (green) and $E_y$ (blue), as well as the projection of $E_x-E_y$ (gray).}
\end{figure}

The fact that strong harmonics can be generated by a circularly polarized driver is in contrast to atomic gas harmonics, where a single circularly polarized laser cannot generate harmonics. To better illustrate the HHG in graphene driven by circularly polarized laser, we plot the harmonic spectra in Fig.~\ref{cp}(a). Up to 13th harmonic can be observed. An essential feature is that only the harmonic orders of $n = 6m \pm 1$ ($m = 1,2,3,\cdots$) are present, while other harmonic orders are all suppressed. This interesting feature looks similar to the cases of HHG driven by two-color counter-rotating circularly polarized laser fields from gases\cite{Kfir2015,Fleischer2014,Pisanty2014} or plasmas\cite{Chen2018}, where spectral selection rules of $n = 3m \pm 1$ can be deduced from simple arguments based on three-fold symmetry and conservation laws for energy, parity, and spin angular momentum\cite{Kfir2015,TD2017b,Fleischer2014,Pisanty2014,Chen2018}. The situation is different here, since a single circularly polarized driver is used. However, the same arguments made by Kfir \textit{et al.} in Ref.~[\citen{Kfir2015}] can be applied to deduce the selection rule here based on symmetry of the interaction system. In Ref.~[\citen{Kfir2015}], the laser field has three-fold rotational symmetry in the polarization plane and the medium is assumed isotropic, and thus the interaction system can be viewed three-fold symmetric. While in the present case, the medium possesses six-fold symmetry and the laser field can be assumed isotropic in the polarization plane, and thus six-fold symmetry is preserved in the interaction system. In fact, all six-fold materials are also three-fold and two-fold symmetric. The three-fold rotational symmetry of the material suppresses every harmonic except orders of $n = 3m \pm 1$, while the centrosymmetry suppresses all even harmonics. These two effects combined lead to the selection rule of $n = 6m \pm 1$. Therefore, the obtained harmonic structure in Fig.~\ref{cp}(a) clearly reflects the six-fold rotational symmetry property of the graphene hexagonal lattice. Thus the HHG spectrum can provide a purely optical method of probing symmetric properties of 2D materials. 

More importantly, the generated high harmonics are also circularly polarized. To demonstrated this, we show the 3D plot of the electric field vector of the 11th and 13th harmonic order in Figs.~\ref{cp}(b) and \ref{cp}(c), respectively. It can be seen clearly that the polarization state of the high harmonics is indeed circular. This is also true for the 5th and 7th harmonics. It also can be seen that the alternative harmonic orders exhibit opposite helicity, similar to the case of HHG driven by bicircular fields\cite{Kfir2015,Chen2018}. Generation of circularly polarized high harmonics extending to the XUV spectral region by a single-color laser pulse in 2D materials may open a door for developing novel ultra-compact photonic devices and new spectroscopy and imaging techniques for investigating chirality-phenomena at nanoscale. Besides, using cycle-level driving laser pulse, it should be possible to generate isolated attosecond XUV pulse with circular polarization in graphene, which can be useful for applications such as probing ultrafast chiral electronic and spin dynamics.

\section{Conclusion}
In summary, we have investigated HHG in graphene driven by in-plane near-infrared laser fields using an \textit{ab initio} approach based on TDDFT. The HHG processes are proven to be in the nonperturbative regime. The calculated spectra show high harmonics extending to the XUV spectral region can be generated in graphene. Spectrogram of the HHG suggests intraband contribution is the dominate mechanism responsible for the HHG in this study. Ellipticity dependence for different harmonic orders is analyzed, which shows HHG is possible driven by a single circularly polarized laser field in graphene. The corresponding harmonic spectra, reflecting the six-fold rotational symmetry of graphene crystal, may offer a purely optical method for probing symmetric properties of 2D materials. Moreover, the corresponding high harmonics are demonstrated to be circularly polarized. The predictions of this study, extending the limit of existing graphene-based HHG sources to the XUV spectral region and adding a new degree of freedom of polarization state, may have impact on many applications such as developing on-chip attoscecond XUV photonic devices at nanoscale, imaging of Berry phase and chirality, and measurement of magnetic and ultrafast spin dynamics in graphene and other 2D materials.

\section*{Funding}
National Natural Science Foundation of China (11705185) and the Presidential Fund of China Academy of Engineering Physics (YZJJLX2017002).

%%%%%%%%%%%%%%%%%%%%%%% References %%%%%%%%%%%%%%%%%%%%%%%%%

%%%%%%%%%% If using BibTeX:
\bibliography{refs}
%\bibliographystyle{osajnl} %the .bst file

%%%%%%%%%% If preparing manually:
% \newcommand{\enquote}[1]{``#1''}

% \bibitem{Zhang:14}
% Y.~Zhang, S.~Qiao, L.~Sun, Q.~W. Shi, W.~Huang, L.~Li, and Z.~Yang,
%   \enquote{Photoinduced active terahertz metamaterials with nanostructured
%   vanadium dioxide film deposited by sol-gel method,}
%   {\protect\JournalTitle{Optics Express}} \textbf{22}, 11070--11078 (2014).

% {Optical Society}, \enquote{{OSA Publishing},}
%   \url{http://www.osapublishing.org}.

% \bibitem{FORSTER2007}
% P.~Forster, V.~Ramaswamy, P.~Artaxo, T.~Bernsten, R.~Betts, D.~Fahey,
%   J.~Haywood, J.~Lean, D.~Lowe, G.~Myhre, J.~Nganga, R.~Prinn, G.~Raga,
%   M.~Schulz, and R.~V. Dorland, \enquote{Changes in atmospheric consituents and
%   in radiative forcing,} in \enquote{Climate Change 2007: The Physical Science
%   Basis. Contribution of Working Group 1 to the Fourth assesment report of
%   Intergovernmental Panel on Climate Change,}  S.~Solomon, D.~Qin, M.~Manning,
%   Z.~Chen, M.~Marquis, K.~B. Averyt, M.~Tignor, and H.~L. Miler, eds.
%   (Cambridge University Press, 2007).

% \end{thebibliography}

\end{document}